\documentclass{JHEP3}
\usepackage{amssymb}

\def \be {\begin{equation}}
\def \ee {\end{equation}}

\def \bea {\begin{eqnarray}}
\def \eea {\end{eqnarray}}
\def \nn {\nonumber}

\def \a {\alpha}
\def \b {\beta}
\def \g {\gamma}
\def \G {\Gamma}
\def \d {\delta}

\def \m {\mu}
\def \n {\nu}
\def \k {\kappa}

\def \s {\sigma}
\def \r {\rho}
\def \o {\omega}
\def \O {\Omega}
\def \th {\theta}
\def \Th {\Theta}

\def \t {\tau}
\def \dag {\dagger}
\def \p {\partial}

\def\bd{\begin{document}}
\def\ed{\end{document}}
\def\nn{\nonumber}
\def\bea{\begin{eqnarray}}
\def\eea{\end{eqnarray}}
\let\bm=\bibitem
\let\la=\label

\def\N{{\cal N}}
\def\sst{\scriptscriptstyle}
\def\thetabar{\bar\theta}
\def\Tr{{\rm Tr}}
\def\one{\mbox{1 \kern-.59em {\rm l}}}

\def\a{\alpha}      \def\da{{\dot\alpha}}
\def\b{\beta}       \def\db{{\dot\beta}}
\def\c{\gamma}  \def\C{\Gamma}  \def\cdt{\dot\gamma}
\def\d{\delta}  \def\D{\Delta}  \def\ddt{\dot\delta}
\def\e{\epsilon}        \def\vare{\varepsilon}
\def\f{\phi}    \def\F{\Phi}    \def\vvf{\f}
\def\h{\eta}
\def\k{\kappa}
\def\l{\lambda} \def\L{\Lambda}
\def\m{\mu} \def\n{\nu}
\def\o{\omega}
\def\P{\Pi}
\def\r{\rho}
\def\s{\sigma}  \def\S{\Sigma}
\def\t{\tau}
\def\th{\theta} \def\Th{\Theta} \def\vth{\vartheta}
\def\X{\Xeta}
\def\z{\zeta}
\def\w{\wedge}
\def\u{\underline}
\def\hs{\hspace}

\def\cA{{\cal A}} \def\cB{{\cal B}} \def\cC{{\cal C}}
\def\cD{{\cal D}} \def\cE{{\cal E}} \def\cF{{\cal F}}
\def\cG{{\cal G}} \def\cH{{\cal H}} \def\cI{{\cal I}}
\def\cJ{{\cal J}} \def\cK{{\cal K}} \def\cL{{\cal L}}
\def\cM{{\cal M}} \def\cN{{\cal N}} \def\cO{{\cal O}}
\def\cP{{\cal P}} \def\cQ{{\cal Q}} \def\cR{{\cal R}}
\def\cS{{\cal S}} \def\cT{{\cal T}} \def\cU{{\cal U}}
\def\cV{{\cal V}} \def\cW{{\cal W}} \def\cX{{\cal X}}
\def\cY{{\cal Y}} \def\cZ{{\cal Z}}
\def\bo {\bar{\o}}

\def\ua{\underline{\alpha}} \def\ubb{\underline{\beta}}
\def\ug{\underline{\gamma}}
\def\ub{\underline{\phantom{\alpha}}\!\!\!\beta}
\def\uc{\underline{\phantom{\alpha}}\!\!\!\gamma}
\def\um{\underline{\mu}} \def\un{\underline{\nu}}
\def\ud{\underline\delta}
\def\ue{\underline\epsilon}
\def\una{\underline a}\def\unA{\underline A}
\def\unb{\underline b}\def\unB{\underline B}
\def\unc{\underline c}\def\unC{\underline C}
\def\und{\underline d}\def\unD{\underline D}
\def\une{\underline e}\def\unE{\underline E}
\def\unf{\underline{\phantom{e}}\!\!\!\! f}\def\unF{\underline F}
\def\unm{\underline m}\def\unM{\underline M}
\def\unn{\underline n}\def\unN{\underline N}
\def\unp{\underline{\phantom{a}}\!\!\! p}\def\unP{\underline P}
\def\unq{\underline{\phantom{a}}\!\!\! q}
\def\unQ{\underline{\phantom{A}}\!\!\!\! Q}
\def\unH{\underline{H}}
\def\ul{\underline}

\def\As {{A \hspace{-6.4pt} \slash}\;}
\def\bs {{b \hspace{-6.4pt} \slash}\;}
\def\Ds {{D \hspace{-6.4pt} \slash}\;}
\def\ds {{\del \hspace{-6.4pt} \slash}\;}
\def\ss {{\s \hspace{-6.4pt} \slash}\;}
\def\ks {{ k \hspace{-6.4pt} \slash}\;}
\def\ps {{p \hspace{-6.4pt} \slash}\;}
\def\pas {{{p_1} \hspace{-6.4pt} \slash}\;}
\def\pbs {{{p_2} \hspace{-6.4pt} \slash}\;}

\def\Fh{\hat{F}}
\def\Vh{\hat{V}}
\def\Xh{\hat{X}}
\def\ah{\hat{a}}
\def\xh{\hat{x}}
\def\yh{\hat{y}}
\def\ph{\hat{p}}
\def\xih{\hat{\xi}}

\def\psit{\tilde{\psi}}
\def\Psit{\tilde{\Psi}}
\def\tht{\tilde{\th}}

\def\At{\tilde{A}}
\def\Qt{\tilde{Q}}
\def\Rt{\tilde{R}}
\def\Nt{\tilde{N}}

\def\at{\tilde{a}}
\def\st{\tilde{s}}
\def\ft{\tilde{f}}
\def\pt{\tilde{p}}
\def\qt{\tilde{q}}
\def\vt{\tilde{v}}
\def\nt{\tilde{n}}

\def\delb{\bar{\partial}}
\def\bz{\bar{z}}
\def\bD{\bar{D}}
\def\bB{\bar{B}}
\def\bo {\bar{\o}}

\def\bk{{\bf k}}
\def\bl{{\bf l}}
\def\bp{{\bf p}}
\def\bq{{\bf q}}
\def\br{{\bf r}}
\def\bx{{\bf x}}
\def\by{{\bf y}}
\def\bR{{\bf R}}
\def\bV{{\bf V}}
\def\bd{\begin{document}}

\def\ed{\end{document}}

\def\d{\delta}\def\D{\Delta}\def\ddt{\dot\delta}

\def\p{\partial} \def\del{\partial}
\def\xx{\times}
\def\uno{\mbox{1 \kern-.59em {\rm l}}}

\def\trp{^{\top}}
\def\inv{^{-1}}
\def\dag{{^{\dagger}}}
\def\pr{\prime}

\def\rar{\rightarrow}
\def\lar{\leftarrow}
\def\lrar{\leftrightarrow}

\def\cw{{\cal W}}
\def\cz{{\cal Z}}
\def\tcm{\tilde{\cal M}}
\def\sgn{{\rm sgn}}
\def\sd {d^{4|4}}
\def\lan{\langle}
\def\ran{\rangle}

\title{On Holographic description of the Kerr-Newman-AdS-dS black holes}
\author{Bin Chen\\
Department of Physics,\\
and State Key Laboratory of Nuclear Physics and Technology,\\
and Center for High Energy Physics,\\
Peking University,\\
Beijing 100871, P.R. China\\
\email{bchen01@pku.edu.cn}}

\author{Jiang Long\\
Department of Physics,\\
Peking University,\\
Beijing 100871, P.R. China\\
\email{longjiang0301@gmail.com}}

\date{\today}

\abstract{In this paper, we study the holographic description of the generic four-dimensional non-extremal Kerr-Newman-AdS-dS black holes. We find that if focusing on the near-horizon region, for the massless scalar scattering in the low-frequency limit, there exists hidden conformal symmetry on the solution
space. Similar to the Kerr case, this suggests that the Kerr-Newman-AdS-dS
black hole is dual to a two-dimensional CFT with central charges
$c_L=c_R=\frac{6a(r_++r_\ast)}{k}$ and temperatures $T_L=\frac{k(r_+^2+r_\ast^2+2a^2)}{4\pi
a\Xi(r_++r_\ast)}, T_R=\frac{k(r_+-r_\ast)}{4\pi a\Xi}$. The macroscopic Bekenstein-Hawking entropy could be recovered from the microscopic counting in dual CFT via the Cardy formula.   Using the Minkowski prescription,
we compute the real-time correlators of the scalar, photon and graviton in near horizon geometry of near extremal  Kerr-AdS-dS black hole. In all these cases, the retarded Green's functions and the corresponding absorption cross sections are in perfect match with CFT prediction. We further discuss the
low-frequency scattering of a charged scalar by a Kerr-Newman-AdS-dS black
hole and find the dual CFT description.
 }

\newpage
\begin{document}

\section{Introduction}\label{sec-intro}

The Kerr/CFT correspondence conjectures that a Kerr black
hole with mass $M$ and angular momentum $J$ is dual to a 2D CFT with central charges $c_L=c_R=12J$ and
temperatures $T_L=M^2/2\pi J, T_R=\sqrt{M^4-J^2}/2\pi J$. This correspondence   was first proposed in \cite{AndyWei} by studying the near-horizon geometry of extreme Kerr black hole (NHEK)\cite{Bardeen:1999px},
 and was then improved by the subsequent study\cite{{Castro:2009jf}, matsuo}, especially on the near-extremal Kerr black holes.
 Support of this conjecture has been
found in  the perfect match of the macroscopic Berenstein-Hawking
entropy of the  black hole with the conformal field theory entropy
computed by the  Cardy formula. See  \cite{Lu:2008jk} for some
further studies  of the Kerr/CFT correspondence as well as
generalizations to other spacetime which contain a warped AdS
structure.

Further support of the correspondence was found in the study of
the superradiant scattering off the extreme Kerr black holes
\cite{Bredberg:2009pv}.
In this case,  the Kerr black hole actually becomes near-extremal, and correspondingly the right-moving
sector of dual CFT is excited \cite{Castro:2009jf}. In the near-horizon limit, the
modes of interest are the ones near the super-radiant bound. It was shown in \cite{Bredberg:2009pv}
that the bulk scattering amplitudes
were in precise agreement with the CFT descriptions whose form are
completely fixed by the conformal invariance.
Similar discussions have been generalized to charged
Kerr-Newman \cite{Hartman:2009nz}, multi-charged
Kerr \cite{Cvetic:2009jn} and higher dimensional near-extremal Kerr
black holes. In all these cases, perfect agreements with the dual CFT
descriptions have been found.

In \cite{ChenChu}, it was shown that the real-time correlators of
various perturbations in near-extremal Kerr(-Newman) black hole
could be computed directly from the bulk, following the Minkowski
prescription proposed in AdS/CFT \cite{Son05} and successfully
used in the warped AdS/CFT correspondence\cite{Chen:2009cg}. It
allowed us  to perform a test directly on the CFT correlators and
the real-time correlators as obtained by holography. The results
are in perfect agreement with the CFT predictions. The similar
prescription\cite{Barnes:2010jp} has been applied to calculate the three-point
functions in the Kerr/CFT correspondence\cite{Becker:2010jj}.

The support of the Kerr/CFT correspondence for generic  non-extremal Kerr black holes
only appeared very recently. In a remarkable
paper \cite{Castro:2010fd} the authors argued that the existence
of conformal invariance in a near horizon geometry is not a
necessary condition, instead the existence of a local conformal
invariance in the solution space of the wave equation for the
propagating field is sufficient to ensure a dual CFT description. The observation
indicates that even though the near-horizon geometry of a generic Kerr black hole could
be far from the AdS or warped AdS spacetime, the local conformal symmetry on the solution space may still
allow us to associate a CFT description to a Kerr black hole.
Both the microscopic entropy counting and the low frequency scalar
scattering amplitude in the  near region support the picture. The similar treatment has
been successfully applied to the higher-dimensional Kerr black hole\cite{Krishnan:2010pv}, the RN
black hole\cite{Chen:2010as}, the Kerr-Newman black hole\cite{Wang:2010qv,Chen:2010xu}. The hidden
conformal symmetry in the solution space also allows us to use the Minkowski prescription to compute the real-time correlators\cite{Chen:2010xu,{Becker:2010dm}}. For other related study on hidden conformal symmetry in the Kerr/CFT correspondence, see \cite{{Li:2010ch},Chen:2010zw,Krishnan:2010df}.

In this paper, we would like to study the holographic description of four-dimensional Kerr-Newman-AdS-dS black holes. The extreme case has been studied in \cite{Hartman:2008pb}, where a dual CFT description of the extreme Kerr-Newman-AdS-dS black holes has been suggested. It would be interesting to see if the picture could be pushed away from the extreme limit. For the generic non-extremal Kerr-Newman-AdS-dS black holes,
the function determining the black hole horizons is quartic, which makes the problem intractable. Nevertheless, we find that in the near-horizon region
there still exists a conformal symmetry acting on the solution space of radial wave function for the massless scalar scattering off the black hole in the low frequency limit.
This is quite different from the treatment on other black holes in the literature. For the low frequency scattering of a Kerr black hole, one may just consider the ``Near" region with $r\o <<1$. It was argued in \cite{Castro:2010fd} that  the hidden conformal symmetry originate from the arbitrariness  in choosing the matching surface between the ``Near" and ``Far" region.
In our case, we have to focus on the near-horizon region, which is much more restricted than the ``Near" region. 
Actually, the scattering problem in the near-horizon region of the Kerr-Newman-AdS-dS black hole is not well-defined. Nevertheless, 
the study of the hidden conformal symmetry of the radial equation  in the near horizon region is still fruitful. The reason originates from
the universal property of the black hole, which suggests that much of the black hole property is captured by the near horizon geometry
of the black hole. The studies of the entropy, the Hawking radiation and the attractor mechanism all support this picture.
In a sense, our study gives another support to this universal picture. 

Firstly we investigate the low-frequency limit of a scalar scattering off a Kerr-AdS-dS black hole. We find that for a
massless scalar, there is a
hidden $SL(2,R)\times SL(2,R)$ conformal symmetry acting on the
solution space of the radial wave function in the near-horizon region. The conformal coordinate transformation allows us to read the corresponding left and right temperatures in the dual CFT:
\be\label{temperature}
T_L=\frac{k(r_+^2+r_\ast^2+2a^2)}{4\pi
a\Xi(r_++r_\ast)}, \hs{3ex}T_R=\frac{k(r_+-r_\ast)}{4\pi a\Xi}.\ee
We calculate the central charge of the dual CFT by studying the near-NHEK geometry and find that
\be\label{centralcharge}
c_L=c_R=\frac{6a(r_++r_\ast)}{k}.\ee
Here $r_+$ is the horizon  and $r_\ast, k$ are parameters depending on the
properties of the black hole. Honestly speaking, these central charges are derived only in the extremal and near-extremal black holes cases. As in the Kerr case, we expect that they still make sense for the generic non-extremal black holes. At the first looking, the above relations are different from the ones found in the literature. But actually they can reduce to the known ones without trouble. In a sense, the above relations is universal, holding in  all cases which have been studied.
As the first nontrivial check, with (\ref{temperature},\ref{centralcharge}), we recover the macroscopic
Bekenstein-Hawking entropy of generic Kerr-Newmann-AdS-dS black holes from the Cardy formula.

Next we use the Minkowski prescription of the AdS/CFT correspondence to calculate the real-time correlators for
the  scalar in the near-horizon geometry of near-extremal Kerr-AdS-dS (near-NHEKS) and find perfect match with  the CFT prediction. The reason we focus on the near-NHEKS is that as
we move away from the near horizon region, the radial equation changes to a much involved form and is hard to solve. Only for near-extremal case, the radial equation is workable.
As a consequence, we have to focus on the frequencies near the super-radiant bound.
Then we turn to the superradiant scattering of vector and
gravitational perturbations off a near-extremal Kerr-AdS-dS black hole. Using the
prescription proposed in \cite{ChenChu}, we compute the real-time
correlators for these perturbations and find perfect
agreement with the CFT prediction.

Finally we study the holographic description of generic Kerr-Newman-AdS-dS black hole. We discuss the scattering
of a charged scalar off the black hole. Once again, in the near-horizon
region, we find the hidden conformal symmetry, which allow us to associate a CFT description of the black hole. We discuss the charged scalar superradiant scattering off a near-extremal Kerr-Newman-AdS-dS black hole, and find that the real-time correlator and greybody factor are in good match with the CFT prediction.

In the next section, we study the low frequency scalar scattering off the Kerr-AdS-dS black hole.
In the near-horizon region, the wave function takes a form of hypergeometric function, suggesting a underlying
conformal invariance.
In section 3, we show the hidden $SL(2,R)\times SL(2,R)$ symmetry acting on the solution space of the massless  scalar wave
function.  In
section 4,  we discuss the microscopic description of  generic non-extremal Kerr-AdS-dS black hole. We obtain the central charges of the near-extremal black holes, and assume that they take the same form for generic black holes. We compute the CFT entropy via the Cardy formula and find it in perfect agreement with the black hole entropy. We also give a brief review of the real-time correlators in 2D CFT.  In section 5,
  we discuss the scattering of scalar, vector and gravitational perturbations off the near-extremal Kerr-AdS-dS
black hole. We compute the real-time correlators from the
Minkowski prescription, and find the agreements with the CFT
predictions. In section 6, we study the holographic description of a  Kerr-Newman-AdS-dS black hole. We end with some discussions in section 7.

\section{Scalar scattering off a Kerr-AdS-dS black hole}

For a four-dimensional Kerr-AdS-dS black hole, its metric takes the following form in Boyer-Lindquist-type coordinates
\be \label{KerrAdS}
ds^2=-\frac{\D_r}{\r^2}\left(d t-\frac{a\sin^2\th}{\Xi} d\phi\right)^2+
\frac{\r^2}{\D_r}dr^2+\frac{\rho^2}{\D_\th} d\th^2+ \frac{\D_\th}{\r^2}\sin^2\th\left(ad
t-\frac{(r^2+a^2)}{\Xi}d\phi\right)^2, \ee where \bea
\Delta_r&=&(r^2+a^2)(1+\frac{r^2}{l^2})-2Mr,  \nn\\
\D_\th &=& 1-\frac{a^2}{l^2}\cos^2 \th, \nn\\
\r^2&=&r^2+a^2\cos^2\th, \nn\\
\Xi &=& 1-\frac{a^2}{l^2}.
\eea
Here $l^{-2}$ is the renormalized cosmological constant, which is positive for dS and negative for AdS.
When $l^{-2}=0$, the above metric reduces to the one of a Kerr black hole.
The physical mass and angular momentum  of the black hole are related to the parameter $M$ by
\be
M_{phy}=\frac{M}{\Xi^2}, \hs{3ex}J=\frac{aM}{\Xi^2}.
\ee

 The black hole horizons are decided by the positive roots of $\D_r=0$, among which the outer one is denoted by $r_+$.  And the Hawking temperature, entropy and
angular velocity of the horizon  are
respectively
 \bea
 T_H&=&\frac{r_+(1+\frac{a^2}{l^2}+\frac{3r^2_+}{l^2}-\frac{a^2}{r^2_+})}{4\pi(r_+^2+a^2)},\nn\\
 S&=&\frac{\pi(r_+^2+a^2)}{\Xi},\nn\\
 \O_H&=&\frac{a\Xi}{r_+^2+a^2}.
 \eea

For simplicity, let us consider a complex massless scalar field  scattering off
the Kerr-AdS-dS black hole. With the ansatz
\be\label{ansatz1}
 \tilde \Phi=e^{-i\o t+im\phi}\Phi,
 \ee
where $\o$ and $m$ are the quantum numbers, the wave equation is of the form
\bea
\lefteqn{\p_r\D_r\p_r\Phi+\frac{(\o(r^2+a^2)-ma\Xi)^2}{\D_r}\Phi+\frac{1}{\sin\th}\p_\th(\D_\th \sin\th\p_\th \Phi)}\nn\\
&&-\frac{(m\Xi)^2}{\D_\th\sin^2\th}\Phi+\left(\frac{2m\o a\Xi}{\D_\th}-\frac{a^2\o^2\sin^2\th}{\D_\th}\right)\Phi =0.
\eea
Let $\Phi= \cR(r)\cS(\th)$, the above equation
could
be decomposed into the angular part and the radial part. The angular
part is of the form \be
 \frac{1}{\sin\th}\frac{d}{d\th}\left(\D_\th\sin\th\frac{d}{d\th}\cS\right)-\frac{(m\Xi)^2}{\D_\th\sin^2\th}\cS+\left(\frac{2m\o a\Xi}{\D_\th}-\frac{a^2\o^2\sin^2\th}{\D_\th}\right)\cS+K\cS=0.
 \ee
Here $K$ is the separation constant.  The radial part of the wave function is of the form
\be
\p_r(\D_r \p_r\cR)+V_R \cR=0
\ee
with
\bea
V_R&=&-K+\frac{(\o(r^2+a^2)-ma\Xi)^2}{\D_r},
\eea

As we are interested in the low frequency limit, $ \o
a <<1$,  the $\o^2$ term in the angular equation could be
neglected. One important consequence is that the separation constant is
independent of $\o$. Another key point in finding the hidden conformal symmetry in the Kerr case is to
focus on the ``Near" region $r \o<<1$, which allows us to simplify the radial equation such that it could be rewritten in terms of the $SL(2,R)$ quadratic Casimir. The same treatment
does not work in the case of Kerr-AdS-dS black hole as the function $\D_r$ is quartic. Nevertheless, we find that if we focus on the near-horizon region then we find the hidden conformal symmetry again. Obviously this region is much restricted than the ``Near" region discussed before. In the near-horizon region, we can expand the function $\D_r$ to the quadratic order of $r-r_+$,
\bea
\D_r&=&(r^2+a^2)(1+\frac{r^2}{l^2})-2Mr, \nn\\
&\simeq & k(r-r_+)(r-r_\ast)
\eea
where $r_+$ is the outer horizon, and
\bea
k&=&1+\frac{a^2}{l^2}+\frac{6r^2_+}{l^2},\label{k}\\
r_\ast&=&r_+-\frac{1}{kr_+}\left(r^2_+-a^2+\frac{3r^4_+}{l^2}+\frac{a^2r^2_+}{l^2}\right).\label{rstar}
\eea
Note that in general $r_\ast$ is not the inner horizon. Only in the cases that $\D_r$ are quadratic, which happens in the Kerr, Kerr-Newman and RN case, $r_\ast$ coincides with the other horizon. Moreover, we have also
\be
r_+^2 \geq \frac{l^2}{6}\left(\sqrt{(1+a^2/l^2)^2+12a^2/l^2}-(1+a^2/l^2)\right),
\ee
in which the equality holds in the extreme case.

As we are in the near-horizon region,
for simplicity, we just focus on the case with also $r_+ \o <<1$. Now the radial equation could be simplified even more
\be\label{scalarKAdS}
\p_r(r-r_+)(r-r_\ast) \p_r\cR(r)+\frac{r_+-r_\ast}{(r-r_+)}A\cR(r)+\frac{r_+-r_\ast}{(r-r_\ast)}B\cR(r)+C\cR(r)=0,\ee
with
\bea
A&=&\frac{(ma\Xi-\o(r^2_++a^2))^2}{k^2(r_+-r_\ast)^2}, \nn\\
B&=&-  \frac{(ma\Xi-\o(r^2_\ast+a^2))^2}{k^2(r_+-r_\ast)^2},\nn\\
C&=&-\frac{K}{k}
\eea
The equation (\ref{scalarKAdS}) has the solution
\be
\cR(z)=z^\a (1-z)^\b F(a,b,c;z)
\ee
with $z=\frac{r-r_+}{r-r_\ast}$ and
\be
\a=-i\sqrt{A},\hs{4ex}\b=\frac{1}{2}(1-\sqrt{1-4C}),
\ee
and
\be
c=1+2\a, \hs{3ex}
a=\a+\b+i\sqrt{-B},\hs{3ex}
b=\a+\b-i\sqrt{-B}. \ee



Note that the radial equation (\ref{scalarKAdS}) holds only at the very near horizon region. However, to discuss the asymptotic behavior of the radial wave function, one has to move away from the near-horizon region, which could make the expansion to the quadratic order problematic. Nevertheless, if one consider the near-horizon region of the near-extremal black holes, one can still apply the same treatment. From the definition $z=\frac{r-r_+}{r-r_\ast}$, only when $r_+\simeq r_\ast$, one do not need to move very far
from the horizon to get $z\to 1$. In this case, one can discuss the scattering amplitudes in the Near-NHEKS geometry, as we will show in section 5.

\section{Hidden conformal symmetry}


In this section, we show that the radial equation (\ref{scalarKAdS}) could be written
in terms of the $SL(2,R)$ quadratic Casimir.

From the conformal coordinates
\bea
\o^+&=&\sqrt{\frac{r-r_+}{r-r_\ast}}e^{2\pi T_R\phi+2n_R t},\nn\\
\o^-&=& \sqrt{\frac{r-r_+}{r-r_\ast}}e^{2\pi T_L\phi+2n_L t},\nn\\
y&=&\sqrt{\frac{r_+-r_\ast}{r-r_\ast}}e^{\pi (T_L+T_R)\phi+(n_L+n_R)t},\nn
\eea
we can locally define the vector fields
\bea
H_1&=&i\p_+ \nn\\
H_0&=&i\left(\o^+\p_++\frac{1}{2}y\p_y\right) \nn\\
H_{-1}&=&i(\o^{+2}\p_++\o^+y\p_y-y^2\p_-)
\eea
and
\bea
\tilde H_1&=&i\p_- \nn\\
\tilde H_0&=&i\left(\o^-\p_-+\frac{1}{2}y\p_y\right) \nn\\
\tilde H_{-1}&=&i(\o^{-2}\p_-+\o^-y\p_y-y^2\p_+)
\eea
These vector fields obey the $SL(2,R)$ Lie algebra
\be
[H_0, H_{\pm 1}]=\mp iH_{\pm 1},\hs{5ex} [H_{-1},H_1]=-2iH_0,
\ee
and similarly for $(\tilde H_0, \tilde H_{\pm 1})$. The quadratic Casimir is
\bea
\cH^2=\tilde{\cH}^2&=&-H_0^2+\frac{1}{2}(H_1 H_{-1}+H_{-1}H_1) \nn\\
 &=&\frac{1}{4}(y^2\p^2_y-y\p_y)+y^2 \p_+\p_-.
 \eea

 In terms of $(t,r,\phi)$ coordinates, the Casimir becomes
 \bea
 \cH^2 &=& (r-r_+)(r-r_\ast)\frac{\p^2}{\p r^2}+(2r-r_+-r_\ast)\frac{\p}{\p r} \\
  &&+\frac{r_+-r_\ast}{r-r_\ast}\left(\frac{n_L-n_R}{4\pi G}\p_\phi -\frac{T_L-T_R}{4G}\p_t\right)^2
  -\frac{r_+-r_\ast}{r-r_+}\left(\frac{n_L+n_R}{4\pi G}\p_\phi
  -\frac{T_L+T_R}{4G}\p_t\right)^2\nn
  \eea
  where $G=n_LT_R-n_R T_L$. We find that
  with the following identification
  \bea
  n_R=0, & & n_L=-\frac{1}{2(r_++r_\ast)} \nn\\
  T_R=\frac{k(r_+-r_\ast)}{4\pi a\Xi}, & & T_L=\frac{k(r^2_++r^2_\ast+2a^2)}{4\pi a\Xi(r_++r_\ast)}, \label{identification}
  \eea
 the radial equation (\ref{scalarKAdS}) of the scalar,  is the same as
 \be
 \tilde{\cH}^2\cR(r)=\cH^2 \cR(r)=K \cR(r).
 \ee
 In other words, the scalar Laplacian is just the $SL(2,R)$ Casimir.

 As pointed out in \cite{Castro:2010fd}, the vector fields are not globally defined. In fact, due to the periodic identification $\phi \sim \phi+2\pi$,
 the $SL(2,R)\times SL(2,R)$ symmetry
 is spontaneously broken down to $U(1)_L\times U(1)_R$ subgroup. As a result, we can
 identify the left and right temperatures in the dual CFT.

 \section{Microscopic description}

From the identification (\ref{identification}), we know the corresponding left and right
temperatures in the dual CFT. In order to have a microscopic description of the black hole, we need to determine the central charges of the dual CFT. For the extremal and near-extremal Kerr black holes, the central charges were derived from the asymptotic symmetry group of the NHEK and near-NHEK geometry\cite{AndyWei, {Castro:2009jf}}, similar to the study of BTZ black hole in three-dimensional gravity\cite{Brown86}.
For a generic non-extremal Kerr
black hole, there is no derivation on the central charges. It was conjectured that the central charges in the near-extremal case could be generalized to the generic cases\cite{Castro:2010fd}. This treatment has been proved to be valid in the Kerr and also Kerr-Newman black holes\cite{Chen:2010xu}. Here we will follow the same logic.

We start from the near-extreme
black hole and consider its near-horizon geometry. Similar to the Kerr case, let us try the following coordinate transformation
\be\label{coordinate}
r=\frac{r_++r_\ast}{2}+\epsilon r_0 \hat r, \hs{3ex} r_+-r_\ast=\l \epsilon r_0, \hs{3ex} t=\hat t \frac{r_0}{\epsilon},\hs{3ex} \phi=\hat \phi+\O_H \hat t\frac{r_0}{\epsilon}
\ee
then we have the near-horizon geometry of near-extremal Kerr-AdS-dS black hole
\be
ds^2=\G(\th)\left(-(\hat r-\frac{\l}{2})(\hat r+\frac{\l}{2})d\hat t^2+
\frac{d\hat r^2}{(\hat r-\frac{\l}{2})(\hat r+\frac{\l}{2})}+\a(\th)d\th^2\right)+\g(\th)(d\hat \phi+\tilde p d\hat t)^2
\ee
where
\be
\G(\th)=\frac{\r^2_+r^2_0}{r_+^2+a^2}, \hs{3ex}\a(\th)=\frac{r^2_++a^2}{\D_\th r^2_0},\hs{3ex}\g(\th)=\frac{\D_\th\sin^2\th(r_+^2+a^2)^2}{\r^2_+r^2_0},\nn
\ee
\be
\tilde p=\frac{ar^2_0\Xi(r_++r_\ast)}{(r^2_++a^2)^2},\hs{3ex}\r_+^2=r^2_++a^2\cos^2\th,\hs{3ex}r^2_0=\frac{r^2_++a^2}{k}.\nn
\ee
The same geometry has been discussed in \cite{Rasmussen:2010xd}.

From the general argument in \cite{Hartman:2008pb}, the central charge should be
\bea
c_L=c_R&=&3p\int_0^\pi d\th \sqrt{\G(\th)\a(\th)\g(\th)}\nn\\
&=&6a\frac{r_++r_\ast}{k}.\label{central}
\eea
This result looks different from the one obtained in the literature. However, note that in the extreme limit, we have
\be
c_L=c_R=6a\frac{r^2_++a^2+\frac{a^2r^2_+}{l^2}+9\frac{r^4_+}{l^2}}{r_+ k^2},
\ee
which is consistent with the result found in \cite{Hartman:2008pb,Rasmussen:2010xd}. Honestly speaking, the central charge (\ref{central}) is 
derived in the near-extremal limit, its robustness for the generic non-extremal black holes needs to be checked. One support to this identification is that it gives correct black hole entropy, as we will show very soon. 

Another subtle issue is on the right central charge. Here we actually assume that the left and right central charges should be same, similar to the Kerr case. One support evidence is that when we take $l^{-2}=0$ limit, the above central charges reduce to the ones in the Kerr case. We find no reason to break the  symmetry between the left- and right-movers on this issue.

  The Kerr/CFT correspondence suggests that the Kerr-AdS-dS black hole is dual to a CFT with central charges (\ref{central})
  at finite temperature $(T_L, T_R)$ given in (\ref{identification}).  This should be true for every value of angular momentum.

\subsection{Thermodynamics}

As a first check of this conjecture in the Kerr-AdS-dS case, we show that the entropy of the black hole could be recovered from dual CFT.
The Cardy formula gives the microscopic entropy
\be
S=\frac{\pi^2}{3}(c_L T_L+c_R T_R).
 \ee
From the central charges (\ref{central}) and the temperatures  (\ref{identification}), we have
\be
S=\frac{\pi(r^2_++a^2)}{\Xi}
\ee
which is in perfect agreement with the macroscopic Bekenstein-Hawking area law for the entropy of the Kerr-AdS-dS black hole.

To determine the conjugate charges, we begin with the first law of
thermodynamics \be  \d S=\frac{\d M-\O \d J}{T_H}=\frac{\d E_L}{T_L}+\frac{\d E_R}{T_R}. \ee  The solution is
\bea
\d E_L&=&\frac{r^2_++r_\ast^2+2a^2}{2a\Xi}\d M, \nn\\
\d E_R&=&\frac{r^2_++r_\ast^2+2a^2}{2a\Xi}\d M-\d J. \eea If
we identify \bea
&&\d M=\o, \hs{3ex}\d J=m,\nn\\
&&\o_L=\frac{r^2_++r_\ast^2+2a^2}{2a\Xi}\o, \hs{3ex}\o_R=\frac{r^2_++r_\ast^2+2a^2}{2a\Xi}\o-m, \label{identification1}  \eea we have \be
\d E_L=\o_L, \hs{3ex}\d E_R=\o_R. \ee

\subsection{Correlators in 2D CFT}

Another support to the Kerr/CFT conjecture is on the study of the scattering amplitudes.
We will focus on the scattering in the near-NHEKS region. Before we analyze the scattering amplitudes, let us give a brief review on the correlators in the dual 2D CFT.

In a 2D conformal field theory(CFT), one can define the two-point
function \be
 G(t^+,t^-)=\langle {\cal O}^\dagger_\phi(t^+,t^-){\cal
 O}_\phi(0)\rangle,
\ee where $t^+,t^-$ are the left and right moving coordinates of
2D worldsheet and  ${\cal O}_\phi$ is the operator corresponding
to the field perturbing the black hole. For an operator of
dimensions $(h_L,h_R)$, charges $(q_L,q_R)$ at temperatures
$(T_L,T_R)$ and chemical potentials $(\m_L, \m_R)$, the two-point
function  is dictated by conformal invariance and takes the form
\cite{Cardy:1984bb}:
 \be \label{G-Mink}
 G(t^+,t^-)\sim (-1)^{h_L+h_R}\left(\frac{\pi T_L}{\sinh(\pi T_L
 t^+)}\right)^{2h_L}\left(\frac{\pi T_R}{\sinh(\pi T_R
 t^-)}\right)^{2h_R}e^{iq_L\m_L t^+ +iq_R\m_R t^-}.
 \ee
The CFT absorption cross
 section could be defined with the two-point functions, following
 Fermi's golden rule:
 \be
 \s_{abs}\sim \int dt^+dt^-
 e^{-i\o_Rt^--i\o_Lt^+}[G(t^+-i\epsilon,
  t^--i\epsilon)-G(t^++i\epsilon, t^-+i\epsilon)]
 \ee
Then after being changed into momentum space, the absorption cross
section is
 \bea\label{2Dsection}
 \s &\sim&
 T_L^{2h_L-1}T_R^{2h_R-1}\sinh\left(\frac{\o_L-q_L\m_L}{2T_L}+\frac{\o_R-q_R\m_R}{2T_R}\right)\nn\\
 && \left|\G\left(h_L+i\frac{\o_L-q_L\m_L}{2\pi T_L}\right)\right|^2\left|\G\left(h_R+i\frac{\o_R-q_R\O_R}{2\pi
 T_R}\right)\right|^2.
 \eea

The retarded correlator  $G_R (\o_L, \o_R)$ is analytic on the
upper half complex $\o_{L,R}$-plane and its value along the
positive imaginary $\o_{L,R}$-axis gives the Euclidean correlator:
\be \label{GER} G_E(\o_{L,E}, \o_{R,E}) = G_R(i\o_{L,E},
i\o_{R,E}), \quad \o_{L,E} , \o_{R,E} >0. \ee
At finite temperature, $\o_{L,E}$ and $\o_{R,E}$ take discrete
values of the Matsubara frequencies \be \o_{L,E} =  2 \pi m_L T_L,
\quad \o_{R,E} =  2 \pi m_R T_R, \ee where $m_L, m_R$ are integers
for bosonic modes and are half integers for fermionic modes.

In a 2D CFT , the Euclidean correlator $G_E$ is obtained
by a Wick rotation $t^+ \to i \t_L$, $t^- \to i \t_R$, and is determined by the conformal symmetry. At finite
temperature the Euclidean time is taken to have period $2 \pi/T_L,
2 \pi/T_R$ and  via analytic continuation the momentum space Euclidean correlator is given
by  \cite{Maldacena:1997ih}
 \bea \label{GE}
G_E(\o_{L,E}, \o_{R,E}) &\sim& T_L^{2 h_L-1}  T_R^{2 h_R-1} e^{i
\frac{\bo_{L,E}}{2T_L}} e^{i \frac{\bo_{R,E}}{2T_R}}\nn\\
&&\cdot\G(h_L + \frac{\bo_{L,E}}{2 \pi T_L})\G(h_L -
\frac{\bo_{L,E}}{2 \pi T_L}) \G(h_R + \frac{\bo_{R,E}}{2 \pi
T_R})\G(h_R - \frac{\bo_{R,E}}{2 \pi T_R}), \eea where \be
\bo_{L,E}= \o_{L,E} - i q_L \m_L, \quad \bo_{R,E}= \o_{R,E} - i
q_R \m_R. \ee

It is remarkable that since the absorption cross section is closely related to the retarded Green's function. Actually $P_{abs}=Im(G_R)$. It seems that the retarded Green's function encodes more information. However via the spectral theorem, it could be true that the absorption cross section can determine the retarded Green's function uniquely. We should not take the greybody factor and the real-time correlators as the independent check of the Kerr/CFT correspondence. Nevertheless, it would still be valuable to compute the real-time correlators directly in the framework of AdS/CFT correspondence.

\section{Superradiant scattering off Kerr-AdS-dS black hole}

In \cite{ChenChu}, it was shown that in the Kerr/CFT correspondence, one
can compute the real-time correlators from usual prescription in AdS/CFT correspondence.
In \cite{Chen:2010xu}, we showed that this is still true for the low-frequency scattering off generic non-extremal Kerr(-Newman) black holes. Here we apply the same prescription to compute the retarded Green's functions in the near-NHEKS geometry.

As we discussed before, to study the scattering problem we have to move away from the
near-horizon region. For a generic non-extremal black hole, the radial wave function becomes intractable. However, for the near-extremal Kerr-AdS-dS black hole, we can still
investigate the scattering in the near-horizon region, similar to the study in the Kerr(-Newman) case\cite{Bredberg:2009pv,Hartman:2009nz,{ChenChu}}.

From the coordinate transformation (\ref{coordinate}), we have
\be
\tilde \Phi=e^{-i\o t+im\phi}\Phi=e^{-i(\o-m\O_H)\hat t\frac{r_0}{\epsilon}+im\hat \phi}\Phi.
\ee
This indicates that we need to focus on the frequencies near the superradiant bound
\be
\o-m\O_H=\hat \o \frac{\epsilon}{r_0},
\ee
with $\hat \o$ being finite.

\subsection{Scalar scattering}

For the massless scalar scatting in the near-NHEKS region, the angular equation of the wave function is of the form
\be
 \frac{1}{\sin\th}\frac{d}{d\th}\left(\D_\th\sin\th\frac{d}{d\th}\cS\right)-\frac{(m\Xi)^2}{\D_\th\sin^2\th}\cS+
 \frac{m^2 a\Xi \O_H}{\D_\th}\left(2-\frac{\sin^2\th}{r^2_++a^2}\right)\cS+\hat K\cS=0.
 \ee
 The separation constant $\hat K$ is different from the one in the low-frequency limit. Therefore, the conformal weight of the fields change correspondingly.

For the radial part of the wave function, the solution is just
\be
\cR(z)=z^{\hat \a} (1-z)^{\hat \b} F(\hat a,\hat b,\hat c;z)
\ee
with $z=\frac{\hat r-\l/2}{\hat r+\l/2}$ and
\be
\hat \a=-i\sqrt{\hat A},\hs{4ex}\hat \b=\frac{1}{2}(1-\sqrt{1-4\hat C}),
\ee
and
\bea
&&\hat c=1+2\hat \a, \hs{2ex}\hat a=\hat \a+\hat \b+i\sqrt{-\hat B},\hs{2ex}
\hat b=\hat \a+\hat \b-i\sqrt{-\hat B}, \\
&&\hat A= \frac{\hat \o^2}{\l^2},\hs{3ex}
\hat B= -\left(\frac{\hat \o}{\l}-\frac{2r_+ m\O_H}{k}\right)^2,\hs{3ex}
\hat C=-\frac{\hat K}{k}.\nn \eea

For a scalar field in a black hole background, the prescription for
two-point real-time correlators was first proposed in \cite{Son05}.
It could be simplified as follows. For the scalar wave function
satisfying the ingoing boundary condition at the black hole horizon,
its asymptotic behavior is \be \phi \sim A r^{h-1}+B r^{-h}. \ee
Then taken $A$ as the source term and $B$ as the response term, the
two-point retarded correlator is just \bea G_R &\sim & \frac{B}{A}.
\eea

For the scalar in the near-NHEKS geometry,
its retarded Green's function is just
 \bea
G_R&\sim&\frac{\G(1-2\hat h)}{\G(2\hat h-1)}\frac{\G\left(\hat h+i\frac{\hat \o_L}{2\pi
\hat T_L}\right)\G\left(\hat h+i\frac{\hat \o_R}{2\pi \hat T_R}\right)}
 {\G\left(1-\hat h+i\frac{\hat \o_L}{2\pi \hat T_L}\right)\G\left(1-\hat h+i\frac{\hat \o_R}{2\pi
 \hat T_R}\right)}
  \eea
 with the identifications
 \bea
 &&\hat T_L=\frac{k}{4\pi r_+\O_H}, \hs{3ex}\hat T_R=\frac{kr_0}{4\pi a\Xi}\l \epsilon, \label{identnear1}\\
 & &\hat \o_L=m, \hs{3ex}\hat \o_R=\frac{r_0 k}{a\Xi}\left(\hat \o-\frac{\l r_+ m\O_H}{k}\right)\epsilon,\label{identnear2}\\
& & \hat h=\frac{1}{2}(1+\sqrt{1-4\hat C})
\eea
  This
is in good match with the CFT prediction (\ref{GER}).

It is remarkable that the identifications (\ref{identnear1}) are exactly the same as the ones (\ref{identification})(\ref{identification1}), which were
obtained respectively from the conformal coordinate transformation in the low-frequency limit and the first law of thermodynamics for generic non-extremal black hole. This provides another nontrivial evidence to support the Kerr/CFT correspondence for the Kerr-AdS-dS black holes.

Note also that the identifications (\ref{identnear1},\ref{identnear2}) are slightly different from the ones used in \cite{Bredberg:2009pv,Hartman:2009nz,{ChenChu}}, in which
the right temperature was set to be finite in the scaling limit. To accommodate the Kerr/CFT correspondence for generic non-extremal black hole, the right temperatures and the right frequencies are set to be very small.

\subsection{Photons and gravitons scattering}

 To study the perturbations with
nonvanishing spin,  one has to apply the Newman-Penrose formalism
\cite{Newman:1961qr}. For the Kerr-AdS-dS black hole, this problem has been discussed in
\cite{Chambers:1994ap}(see also {Khanal:1983vb}).
It turned out that the equations of motion of the
 perturbations
can be decomposed into the angular part and the radial part. The wave function is of the form
  \be
  \Psi^s=e^{-i\o t+im\phi}\cR^s(r)\cS^s(\th).
  \ee
$\Psi^s$ are related to the electromagnetic field strength and Weyl tensor for spin-1 and spin-2 perturbations.
The angular and radial functions satisfy the Teukolsky master equations. The angular part takes the form:
\be \label{angular}
 \frac{1}{\sin\th}\frac{d}{d\th}\left(\sin\th\D_\th\frac{d}{d\th}\cS^s(\th)\right)
+\left(K^s-sN(\th)-\frac{M(\th)^2}{\D_\th\sin^2\th}\right)\cS^s(\th)=0,
 \ee
 with
 \bea
 M(\th)&=&m\Xi-\o a\sin^2\th+s\cos\th(1+\frac{a^2}{l^2}-\frac{2a^2\cos^2\th}{l^2}),\nn\\
 N(\th)&=&1+\frac{a^2}{l^2}+4a\o\cos\th-\frac{6a^2\cos^2\th}{l^2}.
 \eea
 The radial wave function is
 \be
 \left(\cD_{-s/2}\D_r\cD_{s/2}^\dagger+2(2s-1)i\o r-K^s\right)\cR^s=0,
 \ee
 where
 \bea
  \cD_n&=&\p_r+\frac{iH_r}{\D_r}+n\frac{\D'_r}{\D_r}, \nn\\
  \cD_n^\dagger&=&\p_r-\frac{iH_r}{\D_r}+n\frac{\D'_r}{\D_r}, \nn\\
  H_r&=&am\Xi-(r^2+a^2)\o.
 \eea
  Here $K^s$ is the separation constant,  satisfying
 \be\label{Lam}
 K^s(s)=K^s(-s).
 \ee
For the scattering in the near-NHEKS geometry, we need to consider the frequencies near the superradiant bound. In this limit, the $\o$ dependent terms in the angular equation could be evaluated at the superradiant bound.
As a result the separation constant is independent of $\o$.

In fact, for the spin 2 case, there are extra terms proportional to the cosmological constant in the angular and radial equation. In the angle equation, this term may change the value of separation constant. In the radial equation, such a term is quadratic to $r^2$. However, in the near-horizon region, such a term contribute a constant and may induce a change of the conformal weight. In the following, we will not keep the track of
such terms, since they are not relevant for our analysis.

Similar to the scalar case, we will focus on the near horizon region $r\to r_+$. For the superradiant scattering in the near-NHEKS geometry,  the radial equation reduces to \be
\frac{d}{d\hat r}(\hat r-\frac{\l}{2})(\hat r+\frac{\l}{2})\frac{d\cR^s(\hat r)}{d\hat r}+\frac{\l}{\hat r-\l/2}A^sR^s(\hat r)+\frac{\l}{\hat r+\l/2}B^sR^s(\hat r)
+C^s\cR^s(\hat r)
=0 \ee with \bea
A^s&=&\left(\frac{\hat \o}{\l}-\frac{is}{2}\right)^2,\nn\\
B^s&=&-\left(\frac{\hat \o}{\l}+\frac{is}{2}-\frac{2m\O_Hr_+}{k}\right)^2,\nn\\
C^s&=&s-\frac{K^s}{k}+\frac{i4sm\O_Hr_+}{k}+\left(\frac{2m\O_Hr_+}{k}-is\right)^2.\nn
\eea
The solution satisfying the ingoing boundary condition at the
horizon could be once again written in terms of hypergeometric
function
 \be
 \cR^s=z^{\a_s} (1-z)^{\b_s} F(a_s,b_s,c_s;z),
 \ee
 where $z=\frac{\hat r-\l/2}{\hat r+\l/2}$ and
 \bea
 \a_s&=&-i\frac{\hat \o}{\l}-\frac{s}{2} \nn\\
 \b_s&=& \frac{1}{2}(1-\sqrt{1-4\hat C^s}) \nn\\
 a_s&=&\b_s-s-i\frac{\hat \o_L}{2\pi \hat T_L}
 \nn\\
 b_s&=&\b_s-i\frac{\hat \o_R}{2\pi \hat T_R}\nn\\
 c_s&=&1-s-i\frac{2\hat \o}{\l}.
 \eea
 Here $\hat C^s=C^s(\o_s)$, where $\o_s=m\O_H$ is the frequency saturating the superradiant bound.

 The asymptotic behavior of the radial wave function is
 \be
 \cR^s(r) \sim A_1^s \hat r^{h_s-1}+A_2^s \hat r^{-h_s},
 \ee
 where
 \bea
 h^s&=&\frac{1}{2}(1+\sqrt{1-4\hat C^s})\nn\\
 A_1^s&=&\frac{\G(2h_s-1)}{\G(-s+h^s-i\frac{\hat \o_L}{2\pi\hat T_L})\G\left(h^s-i\frac{\hat \o_R}{2\pi \hat T_R}\right)}\nn\\
 A_2^s&=&\frac{\G(1-2h_s)}{\G(-s+1-h^s-i\frac{\hat \o_L}{2\pi\hat T_L})\G\left(1-h^s-i\frac{\hat \o_R}{2\pi \hat T_R}\right)}\nn
\eea One may calculate the absorption cross sections following the
way in \cite{Hartman:2009nz} and compare the results with the CFT
prediction. It turns out to be in perfect match. We will not present
the details here. Instead,  we  give an alternative derivation from
the retarded Green's functions.

For the vector and gravitational perturbations, the prescription has
been proposed in \cite{ChenChu}. If the radial wave function of the
perturbation with the spin $s$ satisfying the ingoing boundary
condition at the black hole horizon has asymptotic behavior as \be
\cR^s (r) \sim A_1^s r^{h-1}+A_2^s r^{-h}, \ee then the retarded
Green's function could be \be\label{pres}
 G_R^s \sim \frac{A_2^{-s}}{A_1^{s}}.
 \ee
 In our case, this leads to
 \be
 G^s_R \sim \frac{\G(1-2h^s)}{\G(2h^s-1)}\frac{\G\left(-s+h^s-i\frac{\hat \o_L}{2\pi \hat T_L}\right)\G\left(h^s-i\frac{\hat \o_R}{2\pi \hat T_R}\right)}
 {\G\left(s+1-h^s-i\frac{\hat \o_L}{2\pi
 \hat T_L}\right)\G\left(1-h^s-i\frac{\hat \o_R}{2\pi \hat T_R}\right)},
  \ee
  where $\hat T_{L,R} \hat \o_{L,R}$ are the ones suggested in (\ref{identnear1},\ref{identnear2}).
  Note that in Kerr-AdS-dS case, the chemical potentials $\mu_{L,R}$ are absent.
With the conformal weights of the fields being identified as
 \be
 h^s_R=h^s, \hs{5ex}h^s_L=h^s_R-s,
 \ee
 the above retarded Green's function agrees precisely, up to a
 normalization factor, with the CFT result (\ref{GE}) at the
 Matsubara frequencies. The cross section can be read directly from
 the above Green's function
 \bea
 \s^s &\sim& {\rm Im}(G^s_R) \nn\\
   &\sim & \frac{1}{(\G(h^s_R-1))^2}\sinh\left(\frac{\hat \o_L}{2\hat T_L}+ \frac{\hat \o_R}{2\hat T_R}\right)\times \nn\\
& &\left|\G\left(h^s_L+i\frac{\hat \o_L}{2\pi \hat T_L}\right)\right|^2\cdot
\left|\G\left(h^s_R+i\frac{\hat \o_R}{2\pi \hat T_R}\right)\right|^2. \eea
They agree with the CFT result.

\section{Holographic description of Kerr-Newman-AdS-dS black hole}

For a four-dimensional Kerr-Newman-AdS-dS black hole, its metric takes the following form in Boyer-Lindquist-type coordinates\cite{Caldarelli:1999xj}
\be \label{KerrNewman}
ds^2=-\frac{\D_r}{\r^2}\left(d t-\frac{a\sin^2\th}{\Xi} d\phi\right)^2+
\frac{\r^2}{\D_r}dr^2+\frac{\rho^2}{\D_\th} d\th^2+ \frac{\D_\th}{\r^2}\sin^2\th\left(ad
t-\frac{(r^2+a^2)}{\Xi}d\phi\right)^2, \ee where \bea
\Delta_r&=&(r^2+a^2)(1+\frac{r^2}{l^2})-2Mr+q^2, \hs{3ex}q^2=q^2_e+q^2_m \nn\\
\D_\th &=& 1-\frac{a^2}{l^2}\cos^2 \th, \nn\\
\r^2&=&r^2+a^2\cos^2\th, \nn\\
\Xi &=& 1-\frac{a^2}{l^2}.
\eea
Here $l^{-2}$ is the renormalized cosmological constant, which is positive for dS and negative for AdS.
When $l^{-2}=0$, the above metric reduces to the one of a Kerr-Newman black hole.
The physical mass, angular momentum and charges of the black hole are related to the parameter $M, q_{e,m}$ by
\be
M_{ADM}=\frac{M}{\Xi^2}, \hs{3ex}J=\frac{aM}{\Xi^2},\hs{3ex}Q_{e,m}=\frac{q_{e,m}}{\Xi}.
\ee

The gauge potential and its field strength are respectively \bea
A&=&-\frac{q_er}{\r^2}(d t-\frac{a\sin^2\th}{\Xi} d\phi)-\frac{q_m\cos\th}{\r^2}(adt-\frac{r^2+a^2}{\Xi}d\phi),\\
F&=&-\frac{1}{\r^4}(q_e(r^2-a^2\cos^2\th)+2q_mra\cos\th)(d t-\frac{a\sin^2\th}{\Xi} d\phi)\wedge dr \nn\\
& &+\frac{1}{\r^4}(q_m(r^2-a^2\cos^2\th)-2q_era\cos\th)\sin\th d\th \wedge (adt-\frac{r^2+a^2}{\Xi}d\phi).
\eea

In the following, for simplicity, we just focus on the case with only electric charge $q_e=q,q_m=0$. The Hawking temperature, entropy and
angular velocity of the horizon  are
respectively
 \bea
 T_H&=&\frac{r_+(1+\frac{a^2}{l^2}+\frac{3r^2_+}{l^2}-\frac{a^2+q^2_e}{r^2_+})}{4\pi(r_+^2+a^2)},\nn\\
 S&=&\frac{\pi(r_+^2+a^2)}{\Xi},\nn\\
 \O_H&=&\frac{a\Xi}{r_+^2+a^2}.
 \eea
The electric potential $\Phi$, measured at infinity with respect to the horizon, is
\be
\Phi_e=A_\m \xi^\m|_{r\to \infty}-A_\m \xi^\m|_{r=r_+}=\frac{q_e r_+}{r^2_++a^2},
\ee
where $\xi=\p_t+\O_H\p_\phi$ is the null generator of the horizon.

For a scalar with charge $e$ and mass $\m$, the Klein-Gordon equation is
\be
(\nabla_\mu+ieA_\mu)(\nabla^\mu+ieA^\mu)\tilde\Phi-\mu^2\tilde\Phi=0.
\ee
In the following, we just focus on the massless case.
With the ansatz
\be\label{ansatz}
 \Phi=e^{-i\o t+im\phi}\cS(\th)\cR(r),
 \ee
where $\o$ and $m$ are the quantum numbers, the wave equation could be decomposed into
an angular part and a radial part. The angular part has the form:
\be
\frac{1}{\sin\th}\p_\th\sin\th\D_\th\p_\th \cS(\th)-\frac{(m\Xi)^2}{\D_\th\sin^2\th}\cS(\th)+\frac{2ma\Xi\o-a^2\o^2\sin^2\th}{\D_\th}\cS(\th)=K_q\cS(\th),
\ee
where $K_q$ is the separation constant.
And the radial wave-function is of the form
\be
\p_r \D_r\p_r \cR+\frac{(H-eqr)^2}{\D_r}R-K_q\cR =0.
\ee

In the near horizon region,  the radial equation could be simplified even more
\be
\p_r(r-r_+)(r-r_\ast) \p_r\cR(r)+\frac{r_+-r_\ast}{(r-r_+)}A^q\cR(r)+\frac{r_+-r_\ast}{(r-r_\ast)}B^q\cR(r)+C^q\cR(r)=0,\label{radial}
\ee
with
\bea
A^q&=&\frac{(\o(r^2_++a^2)-ma\Xi-eqr_+)^2}{k^2(r_+-r_\ast)^2}, \nn\\
B^q&=&-  \frac{(\o(r^2_\ast+a^2)-ma\Xi-eqr_\ast)^2}{k^2(r_+-r_\ast)^2},\nn\\
C^q&=&\frac{e^2q^2}{k^2}-\frac{K_q}{k}
\eea

In fact, there is also a hidden conformal symmetry acting on the solution space. For the neutral scalar, the above radial equation could be rewritten in terms of the $SL(2,R)$ quadratic Casimir. The discussion is the same as the one we did in section 3.
In the end, we have the same temperature identification (\ref{identification}).

Similarly, from the near horizon geometry of the near-extremal Kerr-Newman-AdS-dS black holes, we can read out the central charges of the dual CFT. They turn out to be of the same form as (\ref{central}). With them and the temperatures (\ref{identification}), we recover the Bekenstein-Hawking entropy of the black hole from the microscopic counting via the Cardy formula.

From the first law of thermodynamics, $\d S=\frac{\d M-\O \d J-\Phi \d q}{T_H}=\frac{\d E_L}{T_L}+\frac{\d E_R}{T_R}$, we find that
\be
\d E_L=\o_L-q_L\m_L, \hs{5ex}\d E_R=\o_R-q_R\m_R,
\ee
where we have applied the identification
\bea
&&\d M=\o, \hs{3ex}\d J=m,\hs{3ex}\d Q=e, \nn\\
&&\o_L=\frac{r^2_++r_\ast^2+2a^2}{2a\Xi}\o, \hs{3ex}\o_R=\frac{r^2_++r_\ast^2+2a^2}{2a\Xi}\o-m, \label{identification1q} \\
&&q_L=q_R=\d Q=e, \label{identification2}\\
&&\m_L=\frac{q(r^2_++r_\ast^2+2a^2)}{2a\Xi(r_++r_\ast)},
\hs{3ex}\m_R=\frac{q(r_++r_\ast)}{2a\Xi}.\label{identification3} \eea

As in usual Kerr/CFT correspondence, we can associate a CFT description to a generic non-extremal Kerr-Newman-AdS-dS black hole. The dual CFT has the temperatures $(T_L, T_R)$ as (\ref{identification}), the central charges (\ref{identification1}) and the chemical potential (\ref{identification3}).

To lend more support to the above conjecture, let us study the scattering off the black hole more carefully. We focus on the scattering in the near-horizon geometry of near-extremal Kerr-Newman-Kerr-AdS-dS (near-NHEKNS). As before, we need to focus on the frequencies near the  superradiant bound,
\be
\o \sim \o_s=m\O_H+e\Phi_e.
\ee

In order to study the superradiant scattering in the near-NHEKNS region, it is convenient to introduce the coordinate transformation (\ref{coordinate}). The wave function of the radial equation is then
\be
\cR(z)=z^{\a_q} (1-z)^{\b_q} F(a_q,b_q,c_q;z)
\ee
with $z=\frac{r-\l/2}{r+\l/2}$,
\be
\a_q=-i\sqrt{\hat A^q},\hs{4ex}\b_q=\frac{1}{2}(1-\sqrt{1-4\hat C^q}),
\ee
and
\bea
&&c_q=1+2\a_q,\hs{2ex}
a_q=\a_q+\b_q+i\sqrt{-\hat B^q},\hs{2ex}
b_q=\a_q+\b_q-i\sqrt{-\hat B^q}\nn\\
&&\hat A_q=\frac{\hat \o^2}{\l^2},\hs{3ex}\hat B^q=-\left(\frac{\hat \o}{\l}-\frac{2m\O_H r_+}{k}+\frac{e q}{k}\frac{a^2-r^2_+}{a^2+r^2_+}\right),\hs{2ex}\hat C^q=C^q(\o_s)\eea

The solution behaves asymptotically as
\be\label{chargedscalar} \cR(r) \simeq A^q_1 r^{h_q-1}+A^q_2 r^{-h_q} \ee where
$h_q$ is the conformal weight of the scalar field \be
h_q=1-\b_q=\frac{1}{2}(1+\sqrt{1-4\hat C_q}). \ee
Taking the $A^q_1$ as the source and $A^q_2$ as the response, the retarded
Green's function is just
\bea
G_R&\sim& \frac{A^q_2}{A^q_1}\nn\\
&=&\frac{\G(1-2h_q)}{\G(2h_q-1)}\frac{\G\left(h_q+i\left(\frac{2m\O_H r_+}{k}-\frac{eq}{k}\frac{a^2-r^2_+}{a^2+r^2_+}\right)\right)\G\left(h_q+i\left(2\frac{2\hat\o}{\l}-\frac{2m\O_H r_+}{k}+\frac{eq}{k}\frac{a^2-r^2_+}{a^2+r^2_+}\right)\right)}
 {\G\left(1-h_q+i\left(\frac{2m\O_H r_+}{k}-\frac{eq}{k}\frac{a^2-r^2_+}{a^2+r^2_+}\right)\right)\G\left(1-h_q+i\left(2\frac{2\hat\o}{\l}-\frac{2m\O_H r_+}{k}+\frac{eq}{k}\frac{a^2-r^2_+}{a^2+r^2_+}\right)\right)}\nn\\
&=&\frac{\G(1-2h_q)}{\G(2h_q-1)}\frac{\G\left(h_q+i\frac{\o_L-q_L\m_L}{2\pi
T_L}\right)\G\left(h_q+i\frac{\o_R-q_R\m_R}{2\pi T_R}\right)}
 {\G\left(1-h_q+i\frac{\o_L-q_L\m_L}{2\pi T_L}\right)\G\left(1-h_q+i\frac{\o_R-q_R\m_R}{2\pi
 T_R}\right)},
\eea
with the identification (\ref{identification1q}-\ref{identification3}) and (\ref{identification}). It is  in consistent with the CFT prediction. So is the absorption cross section.

\section{Discussions}

In this paper, we showed that there existed a hidden conformal
symmetry in the low-frequency scattering off the Kerr(-Newman)-AdS-dS black
holes as well. Different from the Kerr or Kerr-Newman case, we had to focus on the near-horizon region, as the function deciding the horizon is quartic. In the near-horizon region, the radial equation of the wave function could be rewritten as the $SL(2,R)$ quadratic Casimir, indicating a hidden conformal symmetry acting on the solution space.
This local conformal symmetry is broken by periodic
identification in the configuration space, which allows us to read out the temperatures of the dual CFT.  Consequently, we would like to suggest that
a generic 4D Kerr-AdS-dS black hole is dual to a 2D CFT with the temperatures
\be
T_L=\frac{k(r_+^2+r_\ast^2+2a^2)}{4\pi
a\Xi(r_++r_\ast)}, \hs{3ex}T_R=\frac{k(r_+-r_\ast)}{4\pi a\Xi},\ee
and the central charges
\be
c_L=c_R=\frac{6a(r_++r_\ast)}{k},\ee
which were derived
 by studying the near-NHEK geometry. Here $r_+$ is the outer horizon of the black hole, while $k$ and $r_\ast$  are  determined by (\ref{k},\ref{rstar}). For the Kerr-Newman-AdS-dS black holes, we have similar holographic picture with addition of chemical potentials to the dual CFT.

We presented the evidence to support this holographic picture. The first evidence is that for a generic black hole, the macroscopic entropy could be recovered from the microscopic counting on the degeneracy in CFT via the Cardy formula. The second evidence is from the study on various kinds of superradiant scattering off the near-extremal black holes.  We found that the real-time correlators and so the absorption cross sections were in perfect match with the CFT prediction, under the identification of the quantum numbers derived from the first law of thermodynamics.

In our study, the hidden conformal symmetry and its corresponding conformal coordinates were investigated in the low-frequency limit. The first law of thermodynamics allows us to identify the frequencies and chemical potential in the CFT dual to a generic black hole. However, when we discussed the scattering off the near-extremal black hole, we focused on the frequencies very near the superradiant bound. Thus the perfect match of various real-time correlators with the CFT predictions provides strong support to the picture that a generic Kerr(-Newman)-AdS-dS black hole has a holographic 2D CFT description.

There is a significant difference between our investigation and the other existing one.
For the black holes studied in this paper, we had to focus on the near-horizon region to
look for the hidden conformal symmetry. It would be interesting to ask why we have such
conformal symmetry, since for generic non-extremal black holes we have no freedom in choosing the matching region.
Actually we can only solve the radial equation in the near horizon region and cannot discuss the scattering issue
as the asymptotic behavior is not well-defined in this case. Only in the near-extremal case, we can zoom in the near-horizon
region and study its scattering amplitude. Nevertheless, the radial equation in the near horizon region has
the hidden conformal symmetry which allows us to read the dual left and right temperatures. This is in spirit in
accordance with the universal property of the black hole, which suggests that the black hole properties such as entropy and
Hawking radiation  are determined merely by its near-horizon geometry\cite{Robinson:2005pd,{Xu:2006tq}}.

There are other kinds of black holes which may have dual CFT description. For example, the higher-dimensional extremal Kerr-AdS-dS black holes have been shown to have CFT descriptions. It would be interesting to apply the treatment developed in this paper to study the holographic description of the generic non-extremal black holes in these cases.

\section*{Acknowledgments}

 The work of BC was partially supported by NSFC Grant
No.10775002,10975005, and NKBRPC (No. 2006CB805905).
BC would like to thank the organizer and the participants of the advanced workshop
``Dark Energy and Fundamental Theory" supported by the Special Fund for Theoretical Physics from the National Natural Science Foundation of China with grant no: 10947203 for stimulating discussions and comments.


\begin{thebibliography}{99}

\bibitem{Bardeen:1999px}
  J.~M.~Bardeen and G.~T.~Horowitz,
  ``The extreme Kerr throat geometry: A vacuum analog of AdS(2) x S(2),''
  Phys.\ Rev.\  D {\bf 60}, 104030 (1999)
  [arXiv:hep-th/9905099].


\bibitem{AndyWei}M. Guica, T. Hartman, W. Song and A. Strominger,
``The Kerr/CFT correspondence,", [arXiv:0809.4266].

\bibitem{matsuo}
  Y.~Matsuo, T.~Tsukioka and C.~M.~Yoo,
  ``Another Realization of Kerr/CFT Correspondence,''
  Nucl.\ Phys.\  B {\bf 825}, 231 (2010)
  [arXiv:0907.0303 [hep-th]].
\\
``Yet Another Realization of Kerr/CFT Correspondence,''
  arXiv:0907.4272 [hep-th].

\bibitem{Castro:2009jf}
  A.~Castro and F.~Larsen,
  ``Near Extremal Kerr Entropy from AdS$_2$ Quantum Gravity,''
  JHEP {\bf 0912}, 037 (2009)
  [arXiv:0908.1121 [hep-th]].


\bibitem{Lu:2008jk}
    H.~Lu, J.~Mei and C.~N.~Pope,
  JHEP {\bf 0904}, 054 (2009)
  [arXiv:0811.2225 [hep-th]].
  T.~Azeyanagi, N.~Ogawa and S.~Terashima,
  JHEP {\bf 0904}, 061 (2009)
  [arXiv:0811.4177 [hep-th]].D.~D.~K.~Chow, M.~Cvetic, H.~Lu and C.~N.~Pope,
  Phys.\ Rev.\  D {\bf 79}, 084018 (2009)
  [arXiv:0812.2918 [hep-th]].
H.~Isono, T.~S.~Tai and W.~Y.~Wen,
  arXiv:0812.4440 [hep-th].
T.~Azeyanagi, N.~Ogawa and S.~Terashima,
  Phys.\ Rev.\  D {\bf 79}, 106009 (2009)
  [arXiv:0812.4883 [hep-th]].J.~J.~Peng and S.~Q.~Wu,
  Phys.\ Lett.\  B {\bf 673}, 216 (2009)
  [arXiv:0901.0311 [hep-th]].F.~Loran and H.~Soltanpanahi,
  Class.\ Quant.\ Grav.\  {\bf 26}, 155019 (2009)
  [arXiv:0901.1595 [hep-th]].
C.~M.~Chen and J.~E.~Wang,
  arXiv:0901.0538 [hep-th].
A.~M.~Ghezelbash,
  JHEP {\bf 0908}, 045 (2009)
  [arXiv:0901.1670 [hep-th]].
H.~Lu, J.~w.~Mei, C.~N.~Pope and J.~F.~Vazquez-Poritz,
  Phys.\ Lett.\  B {\bf 673}, 77 (2009)
  [arXiv:0901.1677 [hep-th]].
G.~Compere, K.~Murata and T.~Nishioka,
  JHEP {\bf 0905}, 077 (2009)
  [arXiv:0902.1001 [hep-th]].
K.~Hotta,
  Phys.\ Rev.\  D {\bf 79}, 104018 (2009)
  [arXiv:0902.3529 [hep-th]].
  D.~Astefanesei and Y.~K.~Srivastava,
  Nucl.\ Phys.\  B {\bf 822}, 283 (2009)
  [arXiv:0902.4033 [hep-th]].
A.~M.~Ghezelbash,
  arXiv:0902.4662 [hep-th].
  C.~Krishnan and S.~Kuperstein,
  Phys.\ Lett.\  B {\bf 677}, 326 (2009)
  [arXiv:0903.2169 [hep-th]].
  T.~Azeyanagi, G.~Compere, N.~Ogawa, Y.~Tachikawa and S.~Terashima,
  Prog.\ Theor.\ Phys.\  {\bf 122}, 355 (2009)
  [arXiv:0903.4176 [hep-th]].
 M.~R.~Garousi and A.~Ghodsi,
  arXiv:0902.4387 [hep-th].
X.~N.~Wu and Y.~Tian,
  Phys.\ Rev.\  D {\bf 80}, 024014 (2009)
  [arXiv:0904.1554 [hep-th]].
  J.~Rasmussen,
  arXiv:0908.0184 [hep-th].
J.~J.~Peng and S.~Q.~Wu,
  Nucl.\ Phys.\  B {\bf 828}, 273 (2010)
  [arXiv:0911.5070 [hep-th]].
    J.~Rasmussen,
  arXiv:0909.2924 [hep-th].
   J.~Rasmussen,
  arXiv:1004.4773 [hep-th].




\bibitem{Bredberg:2009pv}
  I.~Bredberg, T.~Hartman, W.~Song and A.~Strominger,
  ``Black Hole Superradiance From Kerr/CFT,''
  arXiv:0907.3477 [hep-th].







\bibitem{Hartman:2009nz}
  T.~Hartman, W.~Song and A.~Strominger,
  ``Holographic Derivation of Kerr-Newman Scattering Amplitudes for General
  Charge and Spin,''
  arXiv:0908.3909 [hep-th].

\bibitem{Cvetic:2009jn}
  M.~Cvetic and F.~Larsen,
  ``Greybody Factors and Charges in Kerr/CFT,''
  JHEP {\bf 0909}, 088 (2009)
  [arXiv:0908.1136 [hep-th]].

\bibitem{Son05}D.T. Son and A.O. Stariets, ``Minkowski-space
correlators in AdS/CFT correspondence: Recipe and applications,"
JHEP {\bf 0209}, 042 (2002), [hep-th/0205051].

\bibitem{Chen:2009cg}
  B.~Chen, B.~Ning and Z.~b.~Xu,
  ``Real-time correlators in warped AdS/CFT correspondence,''
  arXiv:0911.0167 [hep-th].





\bibitem{ChenChu}
  B.~Chen and C.~S.~Chu,
  ``Real-time correlators in Kerr/CFT correspondence,''
  arXiv:1001.3208 [hep-th].


\bibitem{Becker:2010jj}
  M.~Becker, S.~Cremonini and W.~Schulgin,
  ``Extremal Three-point Correlators in Kerr/CFT,''
  arXiv:1004.1174 [hep-th].

  \bibitem{Barnes:2010jp}
  E.~Barnes, D.~Vaman, C.~Wu and P.~Arnold,
  ``Real-time finite-temperature correlators from AdS/CFT,''
  arXiv:1004.1179 [hep-th].


\bibitem{Castro:2010fd}
  A.~Castro, A.~Maloney and A.~Strominger,
  ``Hidden Conformal Symmetry of the Kerr Black Hole,''
  arXiv:1004.0996 [hep-th].



\bibitem{Krishnan:2010pv}
  C.~Krishnan,
  ``Hidden Conformal Symmetries of Five-Dimensional Black Holes,''
  arXiv:1004.3537 [hep-th].

\bibitem{Chen:2010as}
  C.~M.~Chen and J.~R.~Sun,
  ``Hidden Conformal Symmetry of the Reissner-Nordstrom Black Holes,''
  arXiv:1004.3963 [hep-th].

\bibitem{Wang:2010qv}
  Y.~Q.~Wang and Y.~X.~Liu,
  ``Hidden Conformal Symmetry of the Kerr-Newman Black Hole,''
  arXiv:1004.4661 [hep-th].

\bibitem{Chen:2010xu}
  B.~Chen and J.~Long,
  ``Real-time Correlators and Hidden Conformal Symmetry in Kerr/CFT
  Correspondence,''
  arXiv:1004.5039 [hep-th].

\bibitem{Becker:2010dm}
  M.~Becker, S.~Cremonini and W.~Schulgin,
  ``Correlation Functions and Hidden Conformal Symmetry of Kerr Black Holes,''
  arXiv:1005.3571 [hep-th].

\bibitem{Li:2010ch}
  R.~Li, M.~F.~Li and J.~R.~Ren,
  ``Entropy of Kaluza-Klein Black Hole from Kerr/CFT Correspondence,''
  arXiv:1004.5335 [hep-th].

\bibitem{Chen:2010zw}
  D.~Chen, P.~Wang and H.~Wu,
  ``Hidden conformal symmetry of rotating charged black holes,''
  arXiv:1005.1404 [gr-qc].

\bibitem{Krishnan:2010df}
  C.~Krishnan,
  ``Black Hole Vacua and Rotation,''
  arXiv:1005.1629 [hep-th].

\bibitem{Hartman:2008pb}
  T.~Hartman, K.~Murata, T.~Nishioka and A.~Strominger,
  ``CFT Duals for Extreme Black Holes,''
  JHEP {\bf 0904}, 019 (2009)
  [arXiv:0811.4393 [hep-th]].

\bibitem{Brown86}
J.D. Brown and M. Henneaux,
``Central Charges in the Canonical Realization of Asymptotic
Symmetries: An Example from Three-dimensional Gravity",
Commun. Math. Phys. {\bf 104},207(1986).

\bibitem{Caldarelli:1999xj}
  M.~M.~Caldarelli, G.~Cognola and D.~Klemm,
  ``Thermodynamics of Kerr-Newman-AdS black holes and conformal field
  theories,''
  Class.\ Quant.\ Grav.\  {\bf 17}, 399 (2000)
  [arXiv:hep-th/9908022].

\bibitem{Rasmussen:2010xd}
J.~Rasmussen,
  ``On the CFT duals for near-extremal black holes,''
  arXiv:1005.2255 [hep-th].


\bibitem{Cardy:1984bb}
  J.~L.~Cardy,
  ``Conformal Invariance And Universality in Finite Size Scaling,''
  J.\ Phys.\  A {\bf 17}(1984)L385.

\bibitem{Maldacena:1997ih}
  J.~M.~Maldacena and A.~Strominger,
  ``Universal low-energy dynamics for rotating black holes,''
  Phys.\ Rev.\  D {\bf 56}, 4975 (1997)
  [arXiv:hep-th/9702015].

\bibitem{Newman:1961qr}
  E.~Newman and R.~Penrose,
  ``An Approach to gravitational radiation by a method of spin coefficients,''
  J.\ Math.\ Phys.\  {\bf 3}, 566 (1962).

\bibitem{Khanal:1983vb}
  U.~Khanal,
  ``Rotating Black Hole In Asymptotic De Sitter Space: Perturbation Of The
  Space-Time With Spin Fields,''
  Phys.\ Rev.\  D {\bf 28}, 1291 (1983).

\bibitem{Chambers:1994ap}
  C.~M.~Chambers and I.~G.~Moss,
  ``Stability of the Cauchy horizon in Kerr-de Sitter space-times,''
  Class.\ Quant.\ Grav.\  {\bf 11}, 1035 (1994)
  [arXiv:gr-qc/9404015].

\bibitem{Robinson:2005pd}
  S.~P.~Robinson and F.~Wilczek,
  ``A relationship between Hawking radiation and gravitational anomalies,''
  Phys.\ Rev.\ Lett.\  {\bf 95}, 011303 (2005)
  [arXiv:gr-qc/0502074].\\
S.~Iso, H.~Umetsu and F.~Wilczek,
  ``Hawking radiation from charged black holes via gauge and gravitational
  anomalies,''
  Phys.\ Rev.\ Lett.\  {\bf 96}, 151302 (2006)
  [arXiv:hep-th/0602146].``Anomalies, Hawking radiations and
  regularity
in rotating black holes,''
  Phys.\ Rev.\  D {\bf 74}, 044017 (2006)
  [arXiv:hep-th/0606018].

\bibitem{Xu:2006tq}
  Z.~Xu and B.~Chen,
  ``Hawking radiation from general Kerr-(anti)de Sitter black holes,''
  Phys.\ Rev.\  D {\bf 75}, 024041 (2007) [arXiv:hep-th/0612261].
  
  \end{thebibliography}
   \end{document}